\documentclass{webofc}
\usepackage[varg]{txfonts}   
\usepackage{lineno}
\usepackage{hyperref}
\hypersetup{colorlinks=true, linkcolor=blue, citecolor=blue, urlcolor=blue}

\begin{document}
\title{Advancing ATLAS DCS Data Analysis with a Modern Data Platform}
%
%
\author{\firstname{Luca} \lastname{Canali}\inst{1}\fnsep\thanks{\email{luca.canali@cern.ch}}
\and\firstname{Andrea} \lastname{Formica}\inst{2}\fnsep\thanks{\email{andrea.formica@cern.ch}}
\and\firstname{Michelle Ann} \lastname{Solis}\inst{3}\fnsep\thanks{\email{msolis@arizona.edu}
\\Copyright 2025 CERN for the benefit of the ATLAS Collaboration. CC-BY-4.0 license.}
}

\institute{
    CERN, CH-1211 Geneva 23 (Switzerland)
\and
    IRFU, CEA, Université Paris-Saclay, F-91191 Gif-sur-Yvette (France) 
\and 
    University of Arizona, (USA)
}

\abstract{%
This paper presents a modern and scalable framework for analyzing Detector Control System (DCS) data from the ATLAS experiment at CERN. The DCS data, stored in an Oracle database via the WinCC OA system, is optimized for transactional operations, posing challenges for large-scale analysis across extensive time periods and devices. To address these limitations, we developed a data pipeline using Apache Spark, CERN’s Hadoop service, and the CERN SWAN platform. This framework integrates seamlessly with Python notebooks, providing an accessible and efficient environment for data analysis using industry-standard tools. The approach has proven effective in troubleshooting Data Acquisition (DAQ) links for the ATLAS New Small Wheel (NSW) detector, demonstrating the value of modern data platforms in enabling detector experts to quickly identify and resolve critical issues.}

\maketitle

\section{Introduction}
\label{intro}
The ATLAS Detector Control System (DCS) \cite{atlas_dcs} provides critical information on the detector's availability and performance. Traditionally, these data support online monitoring and troubleshooting, but are rarely used for large-scale analysis due to constraints in data storage and processing. By enabling DCS data to be analyzed alongside other conditions data, this work opens the door to new insights into detector behavior and performance.
Key objectives of this work include:
\begin{itemize}
    \item Exploring insights through large-scale analysis of DCS data.
    \item Correlating DCS data with conditions and other datasets for advanced troubleshooting.
    \item Utilizing a modern, scalable analysis platform with Python APIs and pre-configured tools.
    \item Showing an example of how the platform has been used by a real-life case of troubleshooting Data Acquisition (DAQ) links for the ATLAS New Small Wheel (NSW) detector.
\end{itemize}

\section{Data Storage, Pipeline and Analysis Framework}
\label{data-pipeline}
DCS data, primarily consisting of time-series measurements, are archived in Oracle databases using the WinCC OA system~\cite{winccoa}. Each detector's data are organized within a dedicated database schema, which structures and manages all relevant tables. The \texttt{EVENTHISTORY} table serves as the central repository, storing sensor IDs, timestamps, and measurement values, with new entries added continuously. This table can exceed \(10^9\) rows annually, requiring efficient storage and retrieval strategies. To address this challenge, range partitioning is implemented, dividing the table into smaller, manageable segments based on predefined intervals, such as monthly partitions, to enhance data access and maintenance.

The data import process is implemented as scheduled Apache Spark~\cite{ApacheSpark} jobs, which read data from the source Oracle database and write it into the Hadoop platform~\cite{ApacheHadoop} as Parquet files. Data are extracted from \texttt{ATONR\_ADG}, the read-only replica of the main production database, \texttt{ATONR}. This approach, also used for accessing conditions data~\cite{oracle_conditions}, ensures that the primary database remains unaffected.

The Spark JDBC (Java Database Connectivity) datasource, configured with the Oracle JDBC driver, is used for data transfer. Jobs are scheduled to run daily, with incremental updates performed for the \texttt{EVENTHISTORY} table to maintain synchronization without reprocessing the entire dataset. Smaller and less dynamic tables are fully overwritten during each update cycle, simplifying maintenance and reducing complexity. Data are partitioned by day, with each partition stored as a separate Parquet file. This chronological organization improves storage efficiency and allows for faster query performance by enabling partition pruning during analysis. The import process is repeated for each detector.

Currently, \(3~\mathrm{TB}\) of DCS data have been imported into the Hadoop ecosystem. This represents approximately \(30\%\) of the data available in the Oracle database, covering only records from 2022 onward. This limited scope underscores the significant potential for expansion to include historical datasets. Furthermore, extending the framework to incorporate additional detectors, as required by the ATLAS community, would further enhance its utility and impact.

\subsection{Big Data Platform}
\begin{figure}[ht]
\centering
\includegraphics[width=1.0\textwidth]{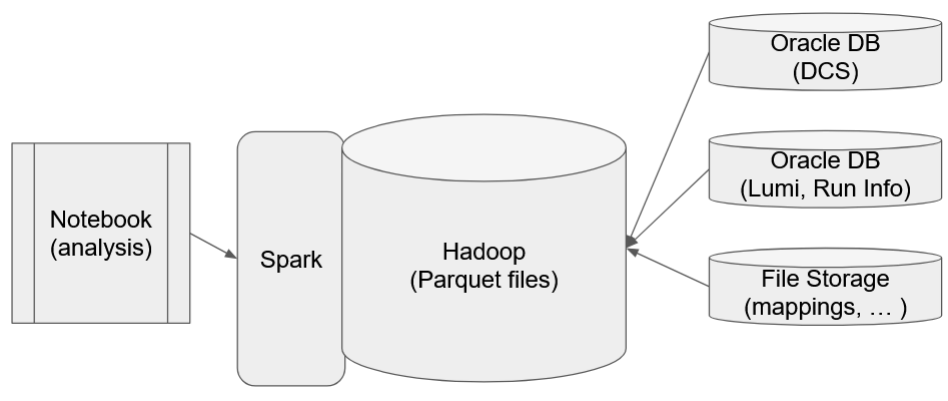}
\caption{Overview of the Big Data architecture for Detector Control System (DCS) data analysis. The system integrates data from Oracle databases (including DCS, luminosity, and run information) and file-based metadata and mappings into the Hadoop ecosystem using Parquet files. Apache Spark serves as the core processing engine, enabling scalable analysis within an interactive environment powered by Jupyter notebooks on CERN SWAN.}
\label{Fig1-bigdata-arch} 
\end{figure}

This project leverages CERN IT services and open-source software to deliver advanced features and high performance while minimizing costs. Figure \ref{Fig1-bigdata-arch} illustrates the components of the Data Platform. CERN's Hadoop cluster, Analytix~\cite{analytix2019}, is a scalable, robust infrastructure with 1400 physical cores and 20 PB of distributed storage, supporting data-intensive projects across CERN.

Apache Hadoop provides a robust distributed computing environment designed for managing large datasets with exceptional reliability and fault tolerance. Its core components, YARN (Yet Another Resource Negotiator) and HDFS (Hadoop Distributed File System), work seamlessly together: YARN dynamically allocates computational resources to enable scalable, concurrent, and high-throughput data processing, while HDFS ensures data durability and fault tolerance by partitioning data into blocks and replicating them across multiple nodes. Together, these components deliver the resource management and storage reliability essential for scaling the analysis of DCS data.

Apache Spark serves as the platform's primary processing engine. Its capability to distribute workloads across the cluster and utilize multiple distributed workers makes it ideal for handling large volumes of data, delivering high performance, scalability, and fault tolerance.

Spark's DataFrame and Spark SQL APIs offer a high-level interface for data manipulation and analysis, streamlining complex workflows. The DataFrame API provides functionality similar to Pandas DataFrames \cite{Pandas}, a popular Python library, while Spark SQL uses SQL, a universal language for querying and managing data across databases and platforms. These APIs are designed to be interchangeable and can be seamlessly combined within Spark, providing exceptional flexibility for querying and processing data. Notably, Spark's Python API, PySpark, enhances accessibility for the physics and data science communities, where Python is a dominant language. 

For data storage in the Hadoop ecosystem, we utilize Apache Parquet~\cite{ApacheParquet}, a modern columnar storage format optimized for analytical workloads. Parquet accelerates data retrieval by enabling selective field access without scanning entire rows, which is ideal for analytical operations. It also offers efficient data compression, reducing storage requirements, and supports predicate pushdown, a feature that filters data at the storage level based on query conditions, reducing the amount of data read and significantly improving query performance.

To support time-based analyses, Parquet files are range-partitioned by timestamps, with daily partitions employed to achieve finer access granularity. This partitioning strategy not only organizes the data logically for efficient retrieval but also enables partition pruning, allowing queries to automatically skip irrelevant partitions based on filter conditions. By minimizing the volume of data scanned during queries, this approach significantly improves query performance, resulting in faster response times and more efficient resource utilization.

By combining Parquet’s compatibility with Spark tools and Python-based workflows, this infrastructure provides a scalable, efficient, and user-friendly environment for offloading, storing, and analyzing Detector Control System (DCS) data at CERN.

\subsection{Analysis Framework}
\label{analysis-framework}

The front-end analysis is performed using Jupyter notebooks, providing an interactive and flexible environment for data exploration and processing. These notebooks integrate seamlessly with Spark, Pandas, and other Python libraries, enabling DCS experts to utilize the data analysis framework efficiently. Key capabilities include:

\begin{itemize}
    \item Querying and processing billions of rows within seconds, utilizing Spark's distributed computation capabilities.
    \item Performing advanced operations, such as data filtering, aggregation, correlation, and feature engineering, with minimal overhead.
    \item Dynamically visualizing data as tables, charts, and plots to uncover trends and detect anomalies in DCS data.
\end{itemize}

The analysis framework is built on CERN's SWAN (Service for Web-based Analysis)~\cite{SWAN}, a Jupyter notebook environment accessible to all CERN users. SWAN enhances accessibility to computational resources by seamlessly integrating with CERN's IT ecosystem, including LCG software releases and CERNBox for file storage and sharing. A standout feature is SWAN's preconfigured Spark connector, which removes the need for intricate user configurations. This connector simplifies the integration of Spark within Jupyter notebooks, enabling efficient and straightforward interaction with CERN's Hadoop service for data analysis.

In addition to working with DCS data, the framework supports querying other data sources, such as Oracle databases and Web Services, using Python libraries. For example, users can retrieve conditions data or detector metadata and combine it with DCS data to conduct comprehensive analyses. This capability greatly enhances the versatility of the platform, making it suitable for both large-scale analytics and metadata integration.

This analysis framework combines flexibility, scalability, and user-friendliness, enabling DCS experts to concentrate on extracting actionable insights without being burdened by infrastructure management.

\subsection{Architecture Summary}
\label{architecture-recap}

The deployed architecture integrates open-source technologies, CERN's IT infrastructure, and DCS data sources into a scalable, high-performance, and user-friendly platform for data analysis. This system empowers detector experts to extract valuable insights from DCS data while providing a solid foundation for future advancements in data-driven detector monitoring and analysis. Key features and components include:

\begin{itemize}
    \item \textbf{Accessibility:} Intuitive Python APIs and Jupyter notebooks on CERN SWAN provide a user-friendly interface, catering to both data scientists and DCS experts.
    \item \textbf{Performance and Scalability:} Apache Spark, integrated with CERN's Hadoop service, drives large-scale data analysis. Efficient data storage and querying are achieved through Parquet's columnar format and time-based partitioning.
    \item \textbf{Versatility:} The architecture integrates seamlessly with multiple data sources, enabling in-depth analyses that extend beyond DCS data alone.
\end{itemize}

To further refine the architecture, several potential improvements and areas of exploration are proposed for future work. These aim to expand its use cases, enhance utility, and ensure long-term adaptability to evolving technological and analytical requirements.

\begin{itemize}
    \item \textbf{Cloud Storage:} Implementing solutions such as Amazon S3 to enhance scalability, flexibility, and cost efficiency for long-term data storage.
    \item \textbf{Kubernetes Orchestration:} Utilizing Kubernetes for Spark orchestration to streamline job deployment, optimize resource allocation, and facilitate a seamless transition to cloud-based infrastructure in case the Hadoop platform becomes unavailable.
    \item \textbf{Frequent Updates:} Incorporating modern data formats like Delta Lake or Apache Iceberg to support more frequent data imports and enable near real-time querying.
    \item \textbf{Machine Learning:} Exploiting GPU resources available on the SWAN platform for advanced analyses, including anomaly detection, predictive maintenance, and correlation studies using deep learning models.
\end{itemize}

\section{ATLAS NSW}
\label{NSW}
This platform provides a ready-to-use solution for querying large volumes of DCS data and quickly beginning analysis. But is it easy to use? Is it helpful? As a use case, we present how the ATLAS NSW benefited from these tools.

\subsection{NSW Introduction}
The ATLAS NSW is part of the muon endcap spectrometer and is a Phase I upgrade, commissioned for Run 3 \cite{nsw-tdr}.  It includes two detectors, the MicroMegas (MMG) and small Thin Gap Chambers (sTGC).  

The NSW DCS monitors critical parameters, including high voltage (HV), low voltage (LV), gas, and electronics. Spanning both wheels, the system comprises 64 sectors, approximately 5000 front-end boards, and over 2 million readout channels. For the MicroMegas (MMG) detector alone, the data volume is substantial, generating 23 million rows of monitoring data daily, amounting to 7.5 billion rows for the year 2024.

\subsection{Key Objectives Driving NSW DCS Data Analysis}
The NSW team has several motivations to perform large-scale analysis with DCS data. 
\begin{itemize}
\item \textbf{Hardware Monitoring:} The VTRx Versatile Transceivers convert optical off-detector communication to electrical communication on the detector.  They were developed at CERN for LHC experiments as a radiation hard upgrade for high-luminosity operation \cite{vtrx-cern}. Failures were observed in these optical links, resulting in communication loss \cite{vtrx-failures}. With more than 1400 VTRx installed on the NSW, it is important to keep track of any potential failures, knowing how many links are affected or have been recovered. They are not easily accessible to replace, but assessing their stability over time could inform how necessary possible interventions would be, for example, during the next long shutdown.  
The Received Signal Strength Indicator (RSSI) current is a parameter in DCS and a diagnostic measurement for the VTRx, proportional to the loss of light.  A high (close to 0.5 V) or unstable RSSI voltage could be a sign of a problematic link.

\item \textbf{DAQ link instability:} The NSW observed DAQ link instability involving approximately 10\% of the VTRx optical fibers.  Several parameters in DCS could be checked for possible correlations, for example, the RSSI. In the end, the RSSI was found not to be correlated and the VTRx was not the cause of this particular DAQ issue, but it was important to rule out an obvious candidate to then focus the investigation of this instability elsewhere.

\item \textbf{Detector performance monitoring:} The HV is crucial for the efficiency of the detector. Using DCS currents and voltages to track the status of HV channels over time, for instance resistive or disabled channels, will be useful for evaluating detector performance, especially over the course of future high-luminosity operation.
\end{itemize}

\begin{figure}[htb]
\centering
\includegraphics[width=1.0\textwidth]{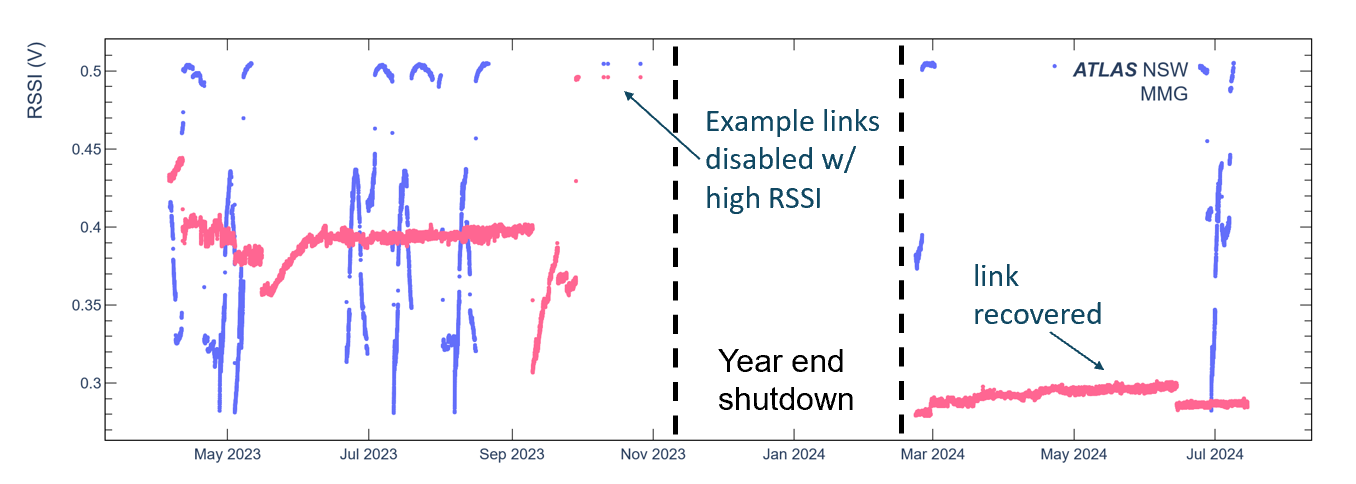}
\caption{Two examples of problematic MMG VTRx optical links over one year, querying DCS RSSI voltage data from over 500 links.  Both links were periodically re-enabled throughout the year to check their status, since the nature of this VTRx failure is unpredictable.  One link remains problematic with the RSSI staying high, while the other link is eventually re-enabled long term, showing a drop in RSSI.}
\label{fig2} 
\end{figure}

\subsection{NSW DCS analysis with the Analytix platform}
The following examples illustrate how this platform was used in studies of the NSW.  Figure \ref{fig2} shows the selection of problematic VTRx over two years, using DCS RSSI data.  An RSSI of 0.5 V is a measure of a link no longer receiving a signal, but it has been observed that links can have communication loss at values less than 0.5 V.  In addition, oscillating RSSI values can indicate a failed link.  Due to the nature of the underlying VTRx issue, where material outgassing can result in a non-uniform degradation of the optical interface, RSSI values jump spontaneously and unpredictably. Equally spontaneous is the possibility of a link recovering on its own.

Since there is no clear RSSI threshold for failures, a query for disabled links was used where the DCS data is no longer updating. During operations, detector experts disable links that stop taking data, from the DAQ and DCS. These links are periodically re-enabled to check their status and disabled again if still problematic. Two such failed links are shown in Figure \ref{fig2}, with periods of being disabled, as well as attempts to re-enable them where the RSSI values remain high. Data is filtered out for the year-end shutdown when the detector is powered off.  Afterwards, one link remains problematic, whereas the other link shows a drop in RSSI, and remains enabled.  This query then gives an accounting of links having problems during data-taking and also shows recovery attempts and their variations in behavior, possibly giving insight into the nature of this issue.

\begin{figure}[htb]
\centering
\includegraphics[width=1.0\textwidth]{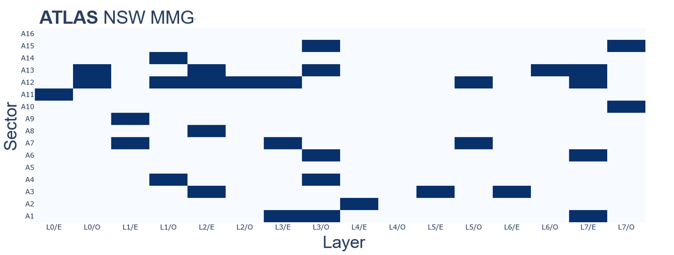}
\caption{MMG VTRx for all 16 sectors of one wheel which exhibited some unstable RSSI behavior.  Each block represents one VTRx, with detector layer number on the x-axis and sector number on the y-axis.  VTRx are selected on the z-axis if RSSI values are over 0.45 V and occurring more than three times, to filter out temporary behaviors, over the course of two years.  Can also select links with a low RSSI standard deviation threshold to identify oscillating values.}
\label{fig3} 
\end{figure}

Regarding platform performance, Spark efficiently queried Parquet files containing over a year of MMG RSSI data from more than 500 optical links using the Spark DataFrame API, completing the query in under 30 seconds. Subsequent data aggregation and reduction tasks, including time binning, geometry selection, and failed VTRx identification, were completed in approximately 3 minutes. For visualization, the data was converted to a Python Pandas DataFrame for plotting.

In Figure \ref{fig3}, the Spark SQL API was used to apply various filters to large datasets, providing a detailed understanding of RSSI behavior. Unstable RSSI values are visualized based on their positions on the detector, facilitating the identification of affected front-end boards and potential correlations with other detector issues. This method also highlights problematic regions within the detector. By caching intermediate DataFrames in Spark, the process of subsequent analysis is streamlined, enabling efficient plotting and visualization of preprocessed data.

\begin{figure}[htb]
\centering
\includegraphics[width=1.0\textwidth]{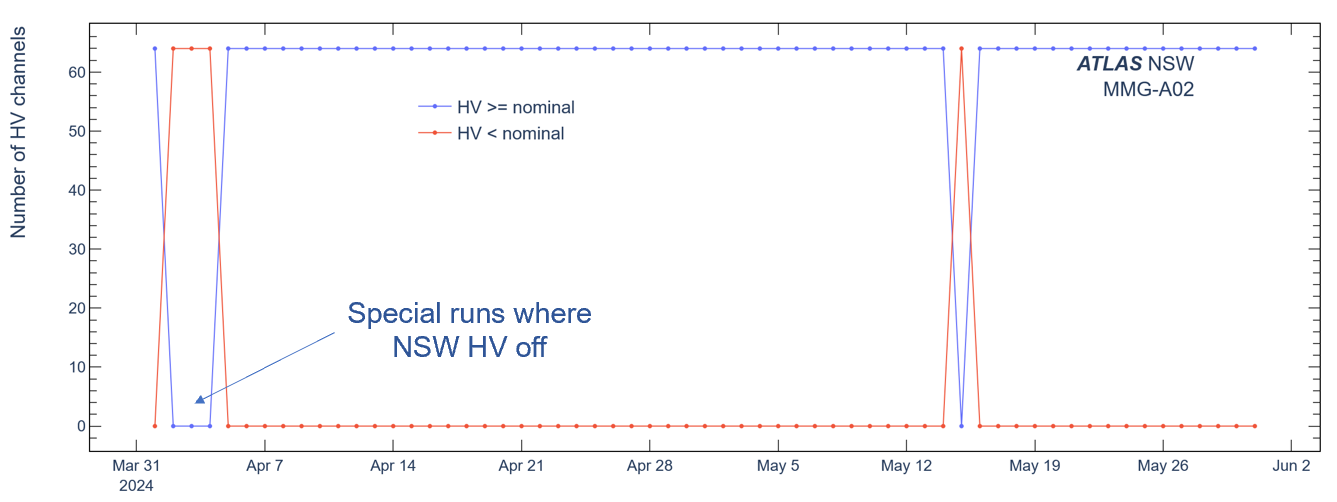}
\caption{Monitored high voltage for the 64 HV channels of one MMG sector. The number of channels with a daily maximum voltage above and below the nominal voltage of 505 V, during data-taking in 2024. Periods where channels are below nominal correspond to special runs where NSW high voltage is off. DCS data joined with Tier 0 processed runs to select times during running in ATLAS.}
\label{fig6} 
\end{figure}

The number of HV channels above and below the nominal monitoring voltage for one sector during data-taking, is presented in Figure \ref{fig6}. We see that all channels for this sector performed at nominal levels except for when the HV was turned off due to special runs. This plot illustrates the ability to quickly join different datasets in Spark, in this case by selecting DCS data and run information from Oracle to select only periods during combined ATLAS running. Since there are known reasons to ramp down the NSW HV outside of data-taking, the HV status during physics would be the most relevant to track.

\section{Conclusions}
\label{sec-crest-conclusions}

This work presents a scalable, efficient, and user-friendly framework for analyzing Detector Control System (DCS) data from the ATLAS experiment at CERN. By leveraging CERN's Hadoop infrastructure, Apache Spark, and Parquet storage, integrated with Jupyter notebooks on the SWAN platform, this methodology addresses the challenges of managing and analyzing large-scale, complex datasets.

The platform has demonstrated its effectiveness in troubleshooting and analyzing detector performance for the ATLAS New Small Wheel (NSW), including resolving DAQ link instabilities and monitoring high-voltage trends. Its flexibility allows for broader applications in detector monitoring and data-driven studies.

This methodology delivers a scalable, accessible framework that advances data-driven analysis for the ATLAS experiment and its collaborators. 

Future enhancements could include replacing Hadoop with Kubernetes for modern, cloud-native orchestration and integrating GPUs to enable machine learning applications, such as anomaly detection and predictive maintenance.

\section*{Acknowledgments}

The authors express their sincere gratitude to the ATLAS ADAM team for their support and to the ATLAS DCS team for their valuable contributions to understanding the DCS data. We also extend special thanks to the ATLAS NSW team for their instrumental collaboration in demonstrating the practical applications of this platform. Finally, we greatly appreciate the CERN IT teams for their exceptional services and support in maintaining the Oracle databases, Hadoop infrastructure, and SWAN platform, which were crucial to this work.

\bibliographystyle{plain}
\bibliography{chep2024-DCS-analysis.bbl}

\end{document}